\documentclass[preprint,showpacs,superscriptaddress,amsmath,amssymb]{revtex4}
\usepackage{dcolumn}
\begin{document}

\preprint{RPC}

\title{Core Polarization with Variable Oscillator Length Parameters \\ for 
Valence Particles}
\author{L. Zamick}
\affiliation{Department of Physics and Astronomy, Rutgers University, 
Piscataway, New Jersey 08854, USA }

\date{\today}

\begin{abstract}
We study quadrupole and monopole core polarization using harmonic oscillator 
wave funtions but with different length parameters for the valence particle as 
compared to the core.  We use perturbation theory with a delta interaction.  
The results also hold for a density dependent delta interaction [1].  We 
study how the amount of core polarization varies with the distance of the 
valence particle from the core.
\end{abstract}

\maketitle

\section{The Quadrupole Mode}

Our simple model consists of a valence p3/2 neutron relative to a $^{4}He$ 
core.  We use a delta interaction to polarize the core.  The particle-hole 
combination excited by the p$_{3/2}$ neutron is 0d 0s$^{-1}$.  We use an 
oscillator length parameter for the cor $b_{c}$.

\begin{equation}
(\hbar w = \hbar^{2} / (2m b^{2}) = 40.46/b^{2})
\end{equation}

We keep $b_{c}$ fixed but we vary the oscillator length parameter b of the valence 
nucleon.

Using scaling properties of harmonic oscillator wave functions and the delta 
interaction we get a simple expression for the ratio of the core polarization 
contribution to the quadrupole moment of $^{5}$He for b(valence) different 
from bc to the case where they are equal.

The ratio for the quadrupole mode is given by

\begin{equation}
RQ(x) = x^{2}/(0.5 + 0.5 x^{2})^{3.5}
\end{equation}
where $x = b/b_{c}$.

Although we expect only small deviations from x=1 it is instructive to study 
the function for all x.

%\begin{center}
\begin{table}
\caption{Quadrupole Core Polarization Ratio vs. x.} 
\begin{tabular}{c|c} 
    x & RQ(x) \\ \hline
 0        &  0  \\  
 0.1  &  0.394  \\  
 $\sqrt{2/5}$  &  1.394  \\ 
 0.9  &  1.149  \\ 
 1 &  1  \\
 1.1 &  0.853 \\ 
 2 & 0.162 \\
 infinity  &  0  \\
\end{tabular}
\end{table}
%\end{center}

Since $x = b/b_{c}$ when x is greater than one, the valence particle is 
further 
away from the core and when x is less than one the valence particle is 
closer to the core.  For values of x close to one we see that the core 
polarization is enhanced when the valence particle moves into the core 
i.e. when the radius of the valence particle orbit is decreased.  Conversely 
when the valence particle moves away from the core it's ability to polarize 
the core is diminished.

In studying the full function with increasing x we see that there is a 
maximum core 
polarization for $x=\sqrt{2/5}$ (RQ=1.3938) after which RQ steadily 
decreases. Near x = 1 RQ(x) is steadily decreasing as the valence particle 
moves away from the core.

\section{The Monopole Mode}

For the monopole mode the particle hole pair is ls 0s$^{-1}$.  Now the ratio 
is

\begin{equation}
RM(x) = (3-2x^{2})/(0.5 + 0.5x^{2})^{3.5}
\end{equation}

\begin{table}
\caption{Monopole Core Polarization Ratio vs. x}
\begin{tabular}{c|c}
x & RM(x) \\ \hline
 0        &  33.941  \\  
 0.9  &  1.957  \\  
 1  &  1 \\ 
 1.1  &  0.409  \\ 
 $\sqrt{3/2}$ &  0  \\ 
 $\sqrt{5/2}$ & -0.282 \\
 infinity  &  0 \\
\end{tabular}
\end{table}

We see that with increasing x, in the monopole case the core polarization 
also decreases especially near x=1.  But then it vanishes when 
$b = \sqrt{3/2}$ $b_{c}$, becomes negative beyond that, reaches an 
extremum at $x = \sqrt{5/2}$ and goes to zero at infinity.

A motivation of this simple model comes from the fact that for neutron rich 
nuclei the valence neutrons can be very loosely bound.  Consequently, their 
ability to polarize the core could be reduced.

\section{References}

[1] J. Speth, L. Zamick and P. Ring, Nucl. Phys. A232, 1 (1974).

\end{document}